\documentclass[11pt]{article}

\usepackage[utf8]{inputenc}
\usepackage[colorlinks=true, allcolors=blue]{hyperref}
\usepackage[margin=1 in]{geometry}
\usepackage{graphicx}
\usepackage{amsmath, amssymb, mathtools, mathrsfs, dsfont, amsthm}
\usepackage{enumerate}
\usepackage{fancyhdr}
\usepackage{authblk}
\usepackage{url}
\usepackage{lipsum}
\usepackage{tensor}
\usepackage{comment}
\graphicspath{/fig}
\usepackage{subcaption}

\newcommand{\dd}{\mathrm{d}}

\theoremstyle{definition}

\theoremstyle{remark}

\theoremstyle{remark}

\title{\textbf{Constraints on cosmologies inside black holes}}
\author{Seamus Fallows\footnote{seamus.fallows@durham.ac.uk} }
\author{Simon F. Ross\footnote{s.f.ross@durham.ac.uk}}
\affil{\textit{Centre for Particle Theory, Department of Mathematical Sciences Durham University, South Road, Durham DH1 3LE, U.K.}}
\date{\today}

\begin{document}
\maketitle
\begin{abstract}
We study the construction of holographic models with closed FRW cosmologies on the worldvolume of a constant-tension brane inside a Schwarzschild-AdS black hole. In dimensions $d>2$, having a smooth Euclidean solution where the brane does not self-intersect limits the brane tension to $T<T_*$, preventing us from realising a separation of scales between the brane and bulk curvature scales. We show that adding interface branes to this model does not relax the condition on the brane tension.   
\end{abstract}

\section{Introduction}

\begin{figure}[ht]
\centering
    \includegraphics[width=.3\linewidth]{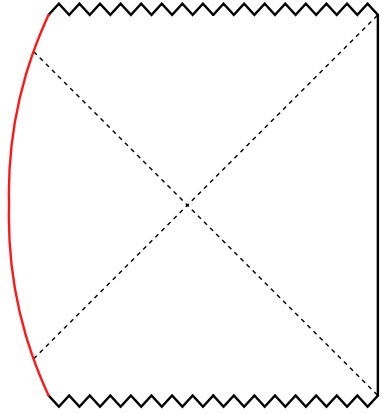}  
\caption{Penrose diagram of an AdS black hole with the left asymptotic region terminating in an ETW brane (shown in red). The worldvolume geometry of the ETW brane is a big-bang/big-crunch cosmology.}
\label{fig1}
\end{figure}

An approach to understanding closed universes with big-bang/big-crunch cosmologies holographically was proposed in \cite{Cooper:2018cmb} (and further developed in \cite{Antonini:2019qkt,VanRaamsdonk:2020tlr,Sully:2020pza,VanRaamsdonk:2021qgv}). The idea is to consider an asymptotically AdS$_{d+1}$ black hole spacetime with a $d$-dimensional dynamical end of the world (ETW) brane behind the horizon providing an inner boundary of the spacetime, as depicted in figure \ref{fig1}. Starting from the $t=0$ surface, the ETW brane falls into the black hole and terminates at the singularity, so its worldvolume geometry is a big-bang/big-crunch cosmology. The state in the bulk on the $t=0$ surface is dual to some state in the dual $d$-dimensional CFT on the asymptotic boundary on the right in figure \ref{fig1}, which therefore includes a description of the cosmology on the ETW brane worldvolume. 

\begin{figure}[ht]
\centering
    \includegraphics[width=\linewidth]{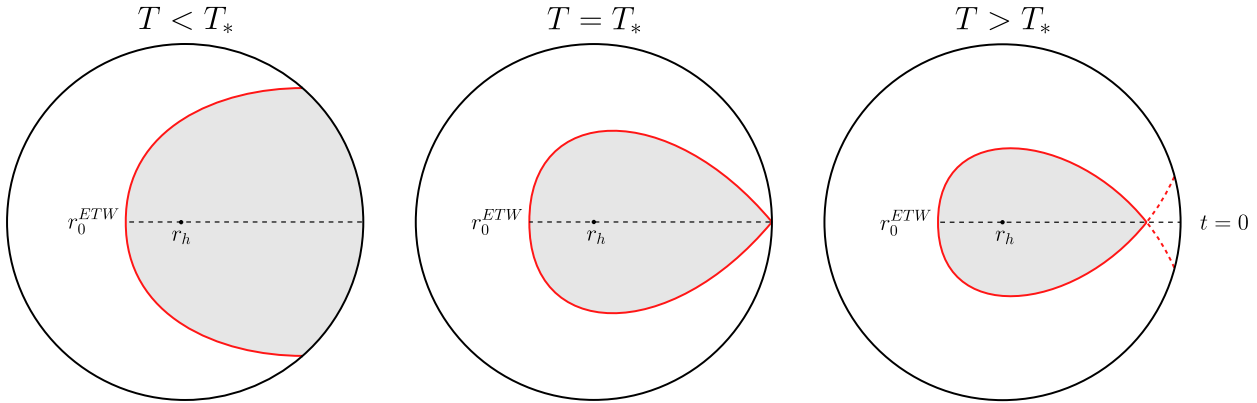}  
\caption{Euclidean ETW brane trajectories for different values of the brane tension. The shaded regions are the bulk regions that we keep from the original Euclidean/Schwarzschild geometry without an ETW brane. Below the critical tension, $T<T_{*}$, there are sensible solutions with the brane intersecting the asymptotic boundary at times $t>0$; at $T=T_{*}$ the brane intersects the asymptotic boundary at $t=0$; for $T>T_{*}$, the brane intersects itself before reaching the asymptotic boundary and we lose the Euclidean path integral construction of the state dual to the $t=0$ slice.}  
\label{fig2}
\end{figure}

An appealing feature of this model is that the state in the bulk on the $t=0$ surface can be constructed from a Euclidean path integral, as depicted on the left in figure \ref{fig2}. In the Euclidean section, the ETW brane moves outward away from $t=0$, and eventually meets the asymptotic boundary. Such solutions were proposed in \cite{Takayanagi:2011zk,Fujita:2011fp} as duals of boundary conformal field theories (BCFTs). The state in the $d$-dimensional CFT dual to the $t=0$ slice in the bulk is then constructed by starting with a $(d-1)$-dimensional boundary state specified by the BCFT and evolving through some period of Euclidean time. Such states have been extensively discussed in recent investigations of black holes, see e.g. \cite{Kourkoulou:2017zaj,Almheiri:2019hni,Penington:2019kki,Chen:2020tes}.

To have a controlled description of the cosmology on the ETW brane, we want to have a separation of scales between the scale that controls the curvature of the ETW brane and the bulk curvature, such that there is a good effective description of the dynamics in terms of ordinary Einstein gravity localised on the ETW brane. This can be achieved by taking the radial position $r_0^{ETW}$ of the brane at $t=0$ to be much larger than the horizon scale $r_h$, which can be achieved by increasing the tension $T$ of the ETW brane \cite{Randall:1999ee,Randall:1999vf,Karch:2000ct}. 

However, in simple examples of this construction this separation of scales is incompatible with the path integral construction described above. If we are in $d>2$, and we take the bulk solution to be an uncharged black hole, increasing $T$ for fixed bulk black hole geometry, there is a critical value $T=T_*$ at which the ETW brane intersects the asymptotic boundary at $t=0$; increasing $T$ beyond $T_*$, the ETW brane will self-intersect before it reaches the asymptotic boundary, as depicted in figure \ref{fig2}. Thus, for $T> T_*$, we lose the Euclidean path integral construction of the state dual to the $t=0$ slice. This would not prevent us from considering the Lorentzian brane solution -- the Lorentzian solution remains well-behaved even for $T> T_*$ -- but we would lose control of the dual CFT state, limiting our ability to study the holographic duality in detail.\footnote{The situation is somewhat similar to considering a small black hole in AdS with a spherical boundary; this is a perfectly good Lorentzian solution, but we do not understand the precise dual CFT description, unlike for large black holes.} One resolution of this problem, proposed in \cite{Antonini:2019qkt}, is to take the bulk solution to be a charged black hole. 

In \cite{VanRaamsdonk:2021qgv}, a different perspective on this issue was given. Suppose the spatial directions along the ETW brane are flat. We can then analytically continue the Euclidean solution along one of the spatial directions, to obtain a time-independent Lorentzian solution where the ETW worldvolume has two asymptotically AdS ends. So the geometry on the ETW brane is an example of an eternal traversable wormhole \cite{Maldacena:2018lmt}. Thus, these solutions can alternatively be viewed as constructing an eternal traversable wormhole by taking the $(d-1)$-dimensional CFTs at the two ends of the ETW brane and coupling them through the $d$-dimensional CFT.\footnote{Similar constructions were considered in \cite{Betzios:2019rds}.} The construction of traversable wormholes requires violations of the energy conditions, and in $d>2$ obstructions to the construction of such wormholes were noted in \cite{Freivogel:2019lej}. In \cite{VanRaamsdonk:2021qgv}, the constraints were analysed from the perspective of the ETW brane worldvolume, and shown to obstruct the construction of traversable wormholes with a separation of scales between the brane and bulk AdS scales. This is the same obstruction as the self-intersection problem mentioned above, as we will explain below. 

\begin{figure}[ht]
\centering
    \includegraphics[width=0.8\linewidth]{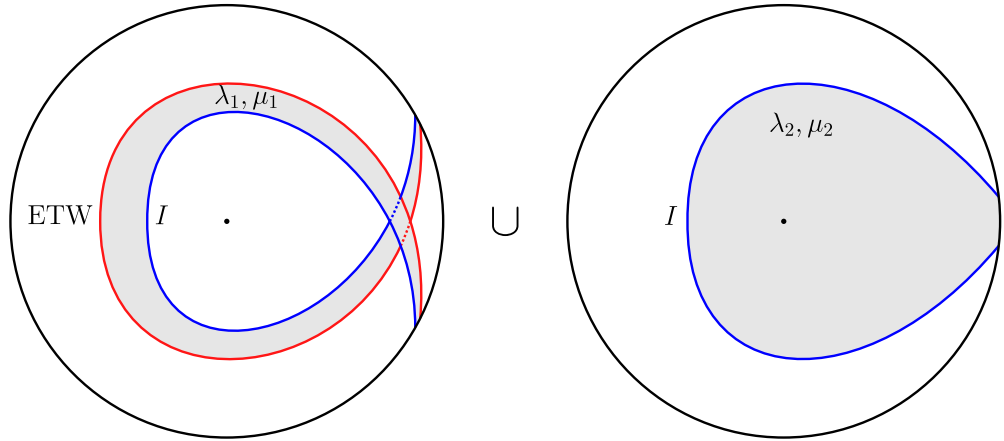}  
\caption{The combined model containing both an ETW brane and an interface brane $I$. The two solutions are glued along the interface brane. Since the Euclidean horizon is not included in the left-hand solution, both branes are multiply wound relative to a single copy of Euclidean/Schwarzschild AdS avoiding self-intersections.}
\label{fig3}
\end{figure}

In \cite{VanRaamsdonk:2021qgv}, it was proposed that this obstruction could be overcome by considering a more complicated bulk solution with interface branes in addition to the ETW brane. From the bulk self-intersection perspective, the advantage of adding interface branes is that one can allow the time at which the ETW brane reaches the boundary to become negative without necessarily encountering a self-intersection, as pictured in figure \ref{fig3} and explained in more detail below. The aim of the present paper is to explore this construction with interface branes in more detail. Unfortunately, we will find that it does not resolve the self-intersection problem; requiring that the ETW brane remains outside of the interface brane and that the interface brane does not self-intersect limits us to $T < T_*$, just as before. Moreover, in the limit as $T$ approaches $T_*$, the interface brane tension must go to zero, returning us to the original setup without an interface brane.

It is not clear if this failure is essentially technical, or is indicative of a deeper issue; the connection to eternal traversable wormholes suggests that it may be the latter. As we will comment on later, the charged black hole solution of \cite{Antonini:2019qkt} provides an example where we obtain a good ETW brane solution and can construct its state by a Euclidean path integral, but this does not define an eternal traversable wormhole, as the analytic continuation that gives us the wormhole makes the Maxwell field associated with the charge imaginary (analytically continuing the charge to make the Maxwell field real would instead break the mechanism that avoids self-intersection). 

The organisation of the rest of the paper is: in the next section, we review the models we consider and relate the worldvolume discussion in \cite{VanRaamsdonk:2021qgv} to the self-intersection picture. In section \ref{const}, we look at the constraints on the model with interface branes. We give a simple scaling argument to show that no solutions without self-intersection exist in the limit of large $r_0^{ETW}$, and explore the regions of the parameter space where solutions exist for finite $r_0^{ETW}$ numerically, showing that solutions exist only for $T<T_*$. In section \ref{disc}, we discuss the differences in the charged black hole case and make some brief concluding comments.

{\it Note added:} after completing this work, we learned that similar results were obtained independently in \cite{Waddell:2022fbn}.  

\section{End of the world brane cosmology}
\label{rev} 

The holographic model of cosmology we consider was first proposed in \cite{Cooper:2018cmb}. The model consists of an AdS black hole bulk with one asymptotic region, with a dynamical constant-tension ETW brane behind the horizon, as pictured in figure \ref{fig1}. The induced geometry on the ETW brane worldvolume is that of a closed FRW universe, with the radial position playing the role of the scale factor. 

Consider first the original setup with just an ETW brane. The bulk action is 
\begin{equation}  
I = \frac{1}{16 \pi G}\left[ \int_\mathcal{M} \dd^{d+1} x \sqrt{-g}\, (R-2\Lambda) + 2\int_{\partial \mathcal{M}}  \dd^d y \sqrt{-h}\, K - 2(d-1) \int_{\mathcal{Q}} \dd^d y \sqrt{-h}\, T \right] ,
\end{equation}
where $\Lambda=-\frac{d(d-1)}{2 L^2}$ is a cosmological constant, $K$ is the trace of the extrinsic curvature and $T$ is the tension of the ETW brane with worldvolume $\mathcal{Q}$, which we take to be one component of the boundary $\partial\mathcal{M}$ of the spacetime, the other component corresponding to the symptotically AdS conformal boundary. 
We consider a planar AdS-Schwarzchild black hole bulk solution
\begin{equation}
    \dd s^2 =  -f(r)\dd t^2+\frac{\dd r^2}{f(r)}+\frac{r^2}{L^2}\dd x^a \dd x_a, \quad \quad f(r)\equiv \frac{r^2}{L^2}-\frac{\mu}{r^{d-2}}.
\end{equation}
This has a horizon at $r=r_h$, where $r_h^d = \mu L^2$. The brane has stress-energy tensor $8\pi G T_{ab}=(1-d)T h_{ab}$, and the action implies that the boundary condition for the bulk metric at $\mathcal{Q}$ is
\begin{equation}
    K_{ab}-Kh_{ab}=(1-d)T h_{ab}
\end{equation}
The $tt$ component of this equation leads to the brane equation of motion 
\begin{equation} \label{braneeom} 
   \left( \frac{\dd r}{\dd t}\right)^2=\frac{f^2(r)}{T^2 r^2}\left(T^2r^2 - f(r)\right).
\end{equation}
In the Lorentzian black hole geometry, the brane will reach a maximum radius $r_0^{ETW}$, with $(r_0^{ETW})^d= \frac{r_h^d}{1 - T^2 L^2}$, which we take to occur at $t=0$. Note $r_0^{ETW} > r_h$ for $T>0$, and $r_0^{ETW} \to \infty$ as $T \to L^{-1}$. To the future and past of this, $r(t)$ decreases, as pictured in figure \ref{fig1}. The brane worldvolume geometry is thus a closed FRW big-bang/big-crunch cosmology, 
where the brane radius $r(t)$ plays the role of the scale factor, and the brane equation of motion \eqref{braneeom} corresponds to the Friedmann equation in this worldvolume cosmology. 

The state on the $t=0$ slice can be obtained by a Euclidean path integral. In the Euclidean black hole, the motion of the ETW brane is 
\begin{equation} \label{Euclidean braneeom} 
   \left( \frac{\dd r}{\dd \tau}\right)^2=\frac{f^2(r)}{T^2 r^2}\left(f(r) - T^2 r^2 \right).
\end{equation}
This now has a minimum at $r=r_0^{ETW}$. Since the ETW brane is inside the black hole in the Lorentzian geometry, it is at its minimum radius in the Euclidean solution at $\tau= \beta/2$, where $\beta = 2\pi L^2/r_h$ is the periodicity in Euclidean time $\tau$. It reaches the AdS boundary at a time  
\begin{equation} \label{tetw}
 \tau^{ETW} = \frac{\beta}{2} - \int_{r_0^{ETW}}^\infty \frac{dr}{f(r)} \frac{Tr}{\sqrt{f(r) - T^2 r^2}}.    
\end{equation}
To avoid self-intersections in the Euclidean solution, we need $\tau^{ETW}>0$. However, setting $r= r_0^{ETW} x$, we have 
\begin{equation}
 \sigma^{ETW} = \frac{2 \tau^{ETW}}{\beta} = 1 - \frac{d}{2\pi} TL (y_0^{ETW})^{\frac{d-2}{2}}  \int_1^\infty \frac{dx}{x^2 (1- (y_0^{ETW})^{-d} x^{-d})  \sqrt{1-x^{-d}}},   
\end{equation}
where $y_0^{ETW} = r_0^{ETW}/r_h$, so $(y_0^{ETW})^{-d} = 1-T^2 L^2$. It is clear that for $d>2$ we can't take $r_0^{ETW} \to \infty$ while keeping $\tau^{ETW} >0$. There must then be some critical value $T= T_* < L^{-1}$ such that $\tau^{ETW} = 0$, and if we consider $T > T_*$ we will have self-intersecting branes in the Euclidean solution.  

It is interesting to note in passing that the integral can be computed exactly: 
\begin{equation}
    \sigma^{ETW} = 1-\frac{d}{2\sqrt{\pi}}TL(1-T^2 L^2)^{\frac{2-d}{2d}}\frac{\Gamma\left(1+\frac{1}{d}\right)}{\Gamma\left(\frac{1}{2}+\frac{1}{d}\right)} {}_2F_1\left(1,\frac{1}{d};\frac{1}{2}+\frac{1}{d};1-T^2 L^2\right).
\end{equation}
For $d=2$, the identity ${}_2F_1(b,a;b;z)= (1-z)^{-a}$ reduces this to $\sigma^{ETW}=\frac{1}{2}$, so the brane hits the boundary at $\tau^{ETW} = \frac{\beta}{4}$ independent of the value of $r_0^{ETW}$, while for $d>2$ the critical tension $T_*$ is defined implicitly by solving $\sigma^{ETW} = 0$.

\subsection{Obstruction from worldvolume perspective}
\label{wv} 

As remarked above, if we take the Euclidean solution and analytically continue one of the flat directions $x^a$, we obtain a solution where the metric on the ETW brane is an eternal traversable wormhole. Thus, the obstruction above can be understood as an obstruction to the existence of such eternal traversable wormholes in the effective induced gravity theory on the brane. 

A general analysis of obstructions from the worldvolume perspective was given in \cite{VanRaamsdonk:2021qgv}, in the spirit of \cite{Freivogel:2019lej}. Following \cite{VanRaamsdonk:2021qgv}, we give the discussion for the case $d=4$. The analysis there is performed in terms of a new coordinate $z$ along the ETW brane, which makes the ETW brane geometry conformally flat, 
\begin{equation}
    \dd s^2 = a^2(z)(\dd z^2+ \eta_{ab} \dd x^a \dd x^b),
\end{equation}
where $a(z)$ has simple poles at $z=\pm z_0/2$ and a minimum at $z=0$.  We assume that we have an effective gravity theory on the brane; the $zz$-component of Einstein's equation then gives
\begin{equation} \label{FRW}
    3\left(\frac{a'}{a}\right)^2-\frac{3a^2}{\ell^2}=8\pi G_{4} T_{zz} ,
\end{equation}
where $G_4$ is the effective 4-dimensional Newton's constant on the brane and $\ell$ is the effective AdS scale on the brane. We take the matter on the brane to be conformal; then $T_{zz}=-3\rho/a^2$. The minimum value of $a$ occurs where $a'=0$ so we have 
\begin{equation}
    a_{min}=(8\pi G_{4} \rho \ell^2)^{\frac{1}{4}}.
\end{equation}
Integrating from this minimum radius to the asymptotically AdS boundary at $z=z_0/2$, we get 
\begin{equation}
    \frac{z_0}{2}= \int_{a_{min}}^{\infty}\frac{\dd a}{\sqrt{\frac{a^4}{\ell^2}-8\pi G_{4}\rho}}=\frac{\ell}{(8 \pi G_4 \rho \ell^2)^{\frac{1}{4}}}I,
\end{equation}
where 
\begin{equation}
    I\equiv \int_{1}^{\infty}\frac{\dd x}{\sqrt{x^4-1}}\approx 1.311.
\end{equation}
Thus we have 
\begin{equation} \label{rhoz} 
    \rho z_0^4 = \frac{2 \ell^2 I^4}{\pi G_{4}} .
\end{equation}
To have an effective gravitational theory on the brane, we want the RHS to be large; we want the brane AdS scale $\ell$ to be large compared to the brane Planck scale. This ratio can also be interpreted as the central charge $c_3$ of the 3d CFT dual to the 4d gravity theory on the brane. But the LHS is naturally of order $c_4$, the number of degrees of freedom of the 4d CFT on the brane. (We get a Casimir energy contribution with $\rho \sim \frac{1}{z_0^4}$ for each species from the finite range of the $z$ coordinate.) 
We cannot solve this problem by taking the number of species $c_4$ parametrically large, of order $c_3$; such a large number of species results in a larger effective UV cutoff on the brane, of order $\ell$ \cite{Freivogel:2019lej}. Thus, there seems to be a general obstruction to the construction of such traversable wormhole solutions. 

For the Schwarzschild-AdS construction considered before, this obstruction is the same as the self-intersection problem.  Using \eqref{braneeom}, the induced metric on the brane in Schwarzschild-AdS is 
\begin{equation}
    \dd s^2= \frac{ f(r)^2}{T^2 r^2} \dd \tau^2+\frac{r^2}{L^2}\dd x^a \dd x_a = \frac{r^2}{L^2}(\dd z^2+\dd x^a \dd x_a),
\end{equation}
so $\dd z^2 = \frac{ f(r)^2 L^2}{T^2 r^4} \dd \tau^2$, and \eqref{braneeom} becomes 
\begin{equation}
    \frac{L^2}{r^2}\left(\frac{\dd r}{\dd z}\right)^2=f(r)-T^2r^2 = r^2 \left(\frac{1}{L^2}-T^2\right) -\frac{\mu}{r^2}.
\end{equation}
This maps to equation (\ref{FRW}) with the identifications 
\begin{equation}
    r^2\to a^2L^2, \quad\quad \left(\frac{1}{L^2}-T^2\right)\to \frac{1}{\ell^2}, \quad\quad \mu\to 8\pi G_4 \rho L^4.
\end{equation}
As stated before, we tune the brane AdS scale $\ell$ to be large by taking the tension $T$ close to $L^{-1}$, and as usual, the energy density on boundary (in this case, a dynamical brane) is set by the black hole mass parameter $\mu$ in the bulk. 

We have $f(r) < \frac{r^2}{L^2}$, so $\dd z < \frac{\dd \tau}{LT}$, and
\begin{equation}
  \rho z_0^4 < \frac{\rho \tau_0^4}{L^4 T^4}   = \left( \frac{\tau_0}{\beta} \right)^4  \frac{L^2}{ G_4} \frac{8\pi^3}{L^4 T^4},
\end{equation}
so we see that $\rho z_0^4$ is indeed naturally of the order of $L^2/G_4 \sim c_4$, the central charge of the CFT dual to the Schwarzschild-AdS bulk, and the problem of making $\rho z_0^4$ large maps on to the problem of making $\tau_0$ large compared to $\beta$, which we can't achieve in this model without self-intersection. 

\subsection{Interface branes}

The proposed solution of this problem is to consider adding an interface brane in the bulk \cite{VanRaamsdonk:2021qgv}. Using results of \cite{Simidzija:2020ukv,Bachas:2021fqo}, \cite{VanRaamsdonk:2021qgv} showed that for extreme values of the interface brane tension, $\rho z_0^4$ could indeed be made large. From the spacetime perspective we will focus on, the idea is that for some choices of the interface tension and bulk parameters, the spacetime on one side of the interface brane doesn't contain a horizon, so both the ETW brane and interface brane can be multiply wound relative to a single copy of the Euclidean Schwarzschild-AdS$_5$ geometry on that side without encountering self-intersections, as pictured in figure \ref{fig3}. We find that this setup doesn't solve the problem however, as the two branes run into each other. In this section we will give a brief overview of interface brane setup; we will discuss the details of the constraints in the next section. 

With an interface brane, we have a spacetime region on either side of the brane, so the action is
\begin{align}
    I = \frac{1}{16 \pi G} & \left[\int_{\mathcal{M}_1}\dd^{d+1}x \sqrt{-g_1}(R_1-2\Lambda_1)+\int_{\mathcal{M}_2}\dd^{d+1}x \sqrt{-g_2}(R_2-2\Lambda_2) \right. \nonumber \\
    & + \left. 2\int_{\mathcal{I}}\dd^{d}y \sqrt{-h}(K_1-K_2)-2(d-1)\int_{\mathcal{I}}\dd^{d}y \sqrt{-h}\kappa - 2(d-1) \int_{\mathcal{Q}} \dd^d y \sqrt{-h}\, T \right],
\end{align}
where $\mathcal{I}\equiv \partial\mathcal{M}_1\cap \partial\mathcal{M}_2$ is the interface brane worldvolume, and $\mathcal Q$ is the worldvolume of the ETW brane as before. We assume a constant tension brane with tension parameter $\kappa$. The motion of the interface brane is determined by the second Israel junction condition 
\begin{equation}
    K_{1ab}-K_{2ab}=\kappa h_{ab}.
\end{equation}
We consider a bulk solution with either side of the interface brane having the metric of a planar AdS black hole of mass $\mu_i$ and AdS scale $L_i\equiv \frac{1}{\sqrt{\lambda_i}}$, 
\begin{equation}
    \dd s_i^2 = f_i(r)\dd t^2+\frac{\dd r^2}{f_i(r)}+r^2\dd x^a \dd x_a, \quad\quad f_i\equiv \lambda_i r^2-\frac{\mu_i}{r^{d-2}}.
\end{equation}
where $i\in\{1,2\}$. The black hole horizon radii are $r_{hi}^d=\frac{\mu_i}{\lambda_i}$. We will have an ETW brane in region 1, so in region 1 $r$ lies in a finite range between the ETW brane and the interface brane, while in region 2 $r$ runs from the interface brane to the asymptotic boundary. The second junction condition leads to 
\begin{equation}
    f_1\frac{dt_1}{ds}-f_2\frac{dt_2}{ds}=\kappa r,
\end{equation}
The first junction condition along with the definition of the proper length parameter $s$ leads to
\begin{equation}
    f_i\left(\frac{dt_i}{ds}\right)+\frac{1}{f_i}\left(\frac{dt_i}{ds}\right)=1.
\end{equation}
Using these we find
\begin{equation}
    \left(\frac{dr}{ds}\right)^2-V_{eff}(r)=0, \quad\quad\quad V_{eff}(r)\equiv \frac{1}{2}(f_1+f_2)-\frac{(f_1-f_2)^2+\kappa^4r^4}{4\kappa^2 r^2},
\end{equation}
and 
\begin{align}
    \frac{dt_1}{dr}&=\frac{1}{f_1\sqrt{V_{eff}(r)}}\left(\frac{1}{2\kappa r}(f_1-f_2)+\frac{1}{2}\kappa r\right), \\
    \frac{dt_2}{dr}&=-\frac{1}{f_2\sqrt{V_{eff}(r)}}\left(\frac{1}{2\kappa r}(f_2-f_1)+\frac{1}{2}\kappa r\right).
\end{align}
The minimum radius of the brane, $V_{eff}(r_0^I)=0$, is found to be
\begin{equation}
(r_0^I)^d = \frac{  (\lambda_1 - \lambda_2) (\mu_1 - \mu_2) - \kappa^2 (\mu_1 + \mu_2) - 2 \kappa \sqrt{ \lambda_1 \mu_2^2 + \lambda_2 \mu_1^2  - (\lambda_1 + \lambda_2 - \kappa^2) \mu_1 \mu_2}}{ \kappa^4 - 2 \kappa^2 (\lambda_1+\lambda_2) + (\lambda_1 - \lambda_2)^2 }.
\end{equation}
It will be useful to define $\bar \lambda = \frac{\lambda_2}{\lambda_1}$, $\bar \mu = \frac{\mu_2}{\mu_1}$ and $\bar \kappa = \frac{\kappa}{\sqrt{\lambda_1}}$. In terms of these parameters the minimum radius is given by 
\begin{equation} \label{r0I}
    (r_0^I)^d = r_{h1}^d\left( \frac{  (1 - \bar \lambda) (1 - \bar \mu) - {\bar \kappa}^2 (1 + \bar \mu) - 2 \bar \kappa \sqrt{\bar \mu {\bar \kappa}^2+\left(\bar \mu -\bar \lambda\right)\left(\bar \mu -1\right)}}{ {\bar \kappa}^4 - 2 {\bar \kappa}^2 (1+\bar \lambda) + (1 - \bar \lambda)^2} \right)
\end{equation}
We want solutions that include a horizon in region 2, and don't include the horizon in region 1. The solution will contain a horizon if $\dot{t_2}<0$ near $r=r^{I}_0$ or
\begin{equation}
    f_2(r^{I}_0)-f_1(r^{I}_0)+\kappa^2 
    (r^{I}_0)^2>0.
\end{equation}
This condition becomes 
\begin{equation} \label{hor}
    \frac{(r^I_0)^d}{r_{h1}^d}\left[{\bar \kappa}^2-(1-\bar \lambda)\right]+(1-\bar \mu) >0.
\end{equation}
From (\ref{r0I}), we see that this condition only depends on the three parameters $(\bar \lambda, \bar \mu, \bar \kappa)$. From (\ref{hor}), we find that the $t_2$ region will contain a horizon if 
\begin{equation} \label{t2 hor}
    \bar \kappa > \sqrt{1-\frac{\bar \lambda}{\bar \mu}} \quad\quad \text{or} \quad\quad \bar \mu < \bar \lambda.
\end{equation}
Requiring there is no horizon in the $t_1$ region, $\dot{t_1}(r^{I}_0)<0$, leads to
\begin{equation}
    \frac{(r^I_0)^d}{r_{h1}^d}\left[{\bar \kappa}^2-(\bar \lambda-1)\right]+(\bar \mu-1) >0.
\end{equation}
giving the condition
\begin{equation} \label{t1 hor}
    \bar \kappa < \sqrt{\bar \lambda-\bar \mu} \quad\quad \text{and} \quad\quad \bar \mu < \bar \lambda.
\end{equation}
Therefore, the desired solution requires we satisfy the second inequality in (\ref{t2 hor}), $\bar \mu <\bar \lambda$. From (\ref{t1 hor}) and ${\bar \kappa}_{min}=|\sqrt{\bar \lambda}-1|$ we get the stronger condition 
\begin{equation} \label{cond}
    \bar \lambda > \frac{1}{4}(1+\bar \mu)^2.
\end{equation}

\section{Constraints on interface brane models}
\label{const}

We now want to consider the constraints we need to satisfy to build well-behaved Euclidean solutions in the interface brane models. Assuming we choose parameters so that the event horizon is not included in region 1, we do not need to impose a periodicity condition on the time coordinate in region 1, so we can have $t^{ETW} <0$ without a self-intersection problem. However, in the presence of the interface brane, we have new constraints from requiring the ETW brane to not run into the interface brane: we need the minimum radius larger, $r_0^{ETW} > r_0^I$, and we need the time at which the ETW brane meets the boundary to be larger, $t^{ETW} > t_1^I$. To avoid self-intersection of the interface brane in region $2$, we also need $t_2^I >0$. We will find that we can satisfy these constraints simultaneously only for $T < T_*$. 

These conditions are conveniently analysed by working in terms of the dimensionless combinations 
\begin{align} \label{sigma^ETW}
\sigma^{ETW} &= \frac{2 t^{ETW}}{\beta_1} = 1 - \frac{2}{\beta_1}\int_{r_0^{ETW}}^\infty \frac{\dd r}{f_1(r)} \frac{Tr}{\sqrt{f_1(r) - T^2 r^2}},  \\
\sigma^I_1 &= \frac{2 t^{I}_1}{\beta_1} = 1 + \frac{2}{\beta_1}\int_{r_0^I}^\infty \frac{\dd r}{f_1(r)} \left( \frac{f_1(r)-f_2(r)+\kappa^2 r^2}{\kappa r \sqrt{V_{eff}(r)}}\right), \\
\sigma^I_2 &= \frac{2 t^{I}_2}{\beta_2} = 1 - \frac{2}{\beta_2}\int_{r_0^I}^\infty \frac{\dd r}{f_2(r)}  \left( \frac{f_2(r)-f_1(r)+\kappa^2 r^2}{\kappa r \sqrt{V_{eff}(r)}}\right),
\end{align}
where $\beta_i = \frac{4\pi}{d \lambda_i r_{hi}}$ is the inverse temperature associated to either side of the interface brane.  

Let us first consider the region of parameters where both $r_0^{ETW}$ and $r_0^I$ are large compared to $r_{hi}$. Large $r_0^{ETW}$ is the regime we wanted to reach, where the theory on the ETW brane is approximately $d$-dimensional Einstein gravity, while we will see that large $r_0^I$ is required to satisfy the constraints; this corresponds to the extreme values of the interface brane tension which were found in \cite{VanRaamsdonk:2021qgv} to give large $\rho z_0^4$. In this limit we can give a simple scaling argument that no good Euclidean solution exists. 

Recall that for the ETW brane, we have 
\begin{equation}
\sigma^{ETW} = 1 - \frac{d}{2 \pi}\frac{T}{\sqrt{\lambda_1}}  (y_0^{ETW})^{\frac{d-2}{2}}  \int_1^\infty \frac{\dd x}{x^2 (1-(y_0^{ETW})^{-d} x^{-d}) \sqrt{1 - x^{-d}}}.   
\end{equation}
It is clear that for $d >2$ we can't take $y_0^{ETW} \to \infty$ while keeping $\sigma^{ETW} >0$. The interface brane scenario relaxes the condition on $\sigma^{ETW}$ to $\sigma^{ETW} > \sigma_1^I$ while keeping $\sigma_2^I >0$. Let us write the integrals in terms of the parameters $(\bar{\lambda}, \bar{\mu}, \bar{\kappa})$ and scale out $r_{h1}$, $r=r_{h1}y$,
\begin{equation}
\sigma_1^I = 1 + \frac{d}{4\pi}\int_{y_0^I}^\infty \frac{\dd y}{\bar{f_1}(y)} \left( \frac{\bar{f_1}(y)-\bar{f_2}(y)+\bar{\kappa}^2 y^2}{\bar{\kappa} y \sqrt{\bar{V}_{eff}(y)}}\right),   
\end{equation}
\begin{equation}
\sigma_2^I = 1 - \frac{d\bar{\lambda}}{4\pi}\frac{r_{h2}}{r_{h1}} \int_{y_0^I}^\infty \frac{\dd y}{\bar{f_2}(y)}  \left( \frac{\bar{f_2}(y)-\bar{f_1}(y)+\bar{\kappa}^2 y^2}{\bar{\kappa} y \sqrt{\bar{V}_{eff}(y)}}\right),   
\end{equation}
where $\bar f_1 = y^2 (1- y^{-d})$, $\bar f_2 = \bar \lambda y^2 (1- \frac{\bar \mu}{\bar \lambda} y^{-d})$, 
\begin{equation}
\bar V_{eff} =   - \frac{y^2}{4\bar \kappa^2} \left[ (\bar \kappa^4 - 2 \bar \kappa^2 (1 + \bar \lambda) + (1 - \bar \lambda)^2 ) + \frac{2}{y^d} (\bar \kappa^2 (1 + \bar \mu) -  (1 - \bar \lambda) (1 - \bar \mu) ) + \frac{1}{y^{2d}} (1 - \bar \mu)^2 \right]. 
\end{equation}
To allow $\sigma^{ETW} \to -\infty$, we want to make $\sigma_1^I$ very negative, by making $y_0^I$ large. For $\bar \kappa$ close to $| 1 - \sqrt{\bar \lambda}|$, we have 
\begin{equation}
(y_0^I)^d \approx \frac{ (\bar \mu - \sqrt{\bar \lambda}) (1- \sqrt{\bar \lambda}) + |\bar \mu - \sqrt{\bar \lambda}| |1- \sqrt{\bar \lambda}|}{2 \sqrt{\bar \lambda} [ \bar \kappa^2 - (1 - \sqrt{\bar \lambda})^2]}.  
\end{equation}
Thus, for $\sqrt{\bar \lambda}$ not between $\bar \mu$ and $1$, $y_0^I$ blows up as $\bar \kappa$ approaches its lower limit $| 1 - \sqrt{\bar \lambda}|$. This is the extremal limit of the interface brane tension introduced in \cite{VanRaamsdonk:2021qgv}. 

In this limit, setting $y = y_0^I /w$, we have
\begin{equation}
\bar V_{eff} \approx   \frac{ (\bar \mu - \sqrt{\bar \lambda}) (1 - \sqrt{\bar \lambda}) }{(y_0^I)^{d-2} \bar \kappa^2} \left(\frac{1- w^{d}}{w^2}\right), 
\end{equation}
and $\bar f_1 \approx (y_0^I)^2 w^{-2}$, $\bar f_2 \approx (y_0^I)^2 \bar \lambda w^{-2} $, so 
\begin{equation}
\sigma_1^I \approx 1 - \frac{d}{2 \pi}(y_0^I)^{\frac{d-2}{2}} \frac{ (\sqrt{\bar \lambda}  -1)}{\sqrt{ ( \sqrt{\bar \lambda} -\bar \mu )(\sqrt{\bar \lambda} -1)}} \int_{0}^1   \frac{\dd w}{ \sqrt{1 - w^{d}}}
\end{equation}
and 
\begin{equation}
\sigma_2^I \approx 1 - \frac{d}{2\pi}(\bar \lambda)^{\frac{d-2}{2d}}(\bar \mu)^{\frac{1}{d}} (y_0^I)^{\frac{d-2}{2}} \frac{ (\sqrt{\bar \lambda}  -1)}   {\sqrt{ ( \sqrt{\bar \lambda} -\bar \mu )(\sqrt{\bar \lambda} -1)}} \int_{0}^1   \frac{\dd w}{\sqrt{1 - w^{d}}}. 
\end{equation}
We see from these that $\sigma_1^I <0$ requires $\bar \lambda>1$. Then to make $\sigma_2^I >0$, we need to make $\bar \mu$ small, so that $ (\bar \mu)^{\frac{1}{d}}(y_0^I)^{\frac{d-2}{2}}$ remains finite as $y_0^I \to \infty$. Note these conditions are consistent with the previous condition on $\bar{\lambda}$, $\bar{\mu}$ given in (\ref{cond}). In this limit of small $\bar \mu$, 
\begin{equation}
\sigma_1^I \approx 1 - \frac{d}{2 \pi}(y_0^I)^{\frac{d-2}{2}} \sqrt{ 1- \frac{1}{\sqrt{\bar \lambda}}}  \int_{0}^1   \frac{\dd w}{ \sqrt{1 - w^{d}}}.
\end{equation}

Since $y_0^{ETW} > y_0^I \gg 1$, the ETW brane tension is close to its upper bound $T \approx \sqrt{\lambda_1}$ and we have that
\begin{equation} \label{cfail} 
\sigma^{ETW} - \sigma_1^I  \approx \left[  (y_0^I)^{\frac{d-2}{2}} \sqrt{ 1- \frac{1}{\sqrt{\bar \lambda}}}   - (y_0^{ETW})^{\frac{d-2}{2}}   \right]\frac{d}{2\pi} \int_{0}^1   \frac{\dd w}{\sqrt{1 - w^{d}}} <0.
\end{equation}
The requirement that the ETW brane is initially outside the interface brane implies that it will run into the interface brane before reaching the boundary. 

Thus, we have shown that our constraints cannot all be satisfied in the region where $y_0^{ETW}$ is large. This is disappointing, as this is the limit of interest. From the perspective of the worldsheet analysis of \cite{VanRaamsdonk:2021qgv} reviewed in section \ref{wv}, increasing $y_0^I$ increases $\rho$, but at the same time it requires us to increase $y_0^{ETW}$, which corresponds to increasing $c_3$. The increasing value of $\rho z_0^4$ never quite catches up with the target value. 

However, it looks like the introduction of the interface brane has improved things {\it somewhat}; while \eqref{cfail} is always negative, we can make it close to zero if we take $\bar \lambda$ large and $y_0^I$ close to $y_0^{ETW}$. Surprisingly, we will see in the next subsection that this seeming improvement is misleading; there are no values of $T$ where we have good Euclidean solutions with interface branes where we did not already have good solutions without interface branes. 

\subsection{Numerical investigation}

In this section, we numerically study the region of the parameter space in which the constraints $\sigma^{ETW} > \sigma^I_1$, $\sigma^I_2 >0$ are satisfied. The relevant parameters are the dimensionless combinations $\bar \lambda, \bar \mu$ and the brane tensions $\bar \kappa, \bar T = T/\sqrt{\lambda_1}$. We show that the solutions with interface branes only exist for choices of $T$ that also admit solutions without interface branes. 

Due to the constraint on $\bar{\lambda}$ in (\ref{cond}), we define 
\begin{equation}
    \bar{\lambda}' \equiv \bar{\lambda}-\bar{\lambda}_{min} = \bar{\lambda}-\frac{1}{4}(1+\bar{\mu})^2.
\end{equation}
Plots will be more clear in terms of this shifted variable. The interface tension is constrained to the interval $\bar{\kappa}\in (\bar{\kappa}_{min},\bar{\kappa}_{max})= (|\sqrt{\bar{\lambda}}-1|,\sqrt{\bar{\lambda}-\bar{\mu}})$, so we parameterise it as
\begin{equation}
    \bar{\kappa}=\bar{\kappa}_{min}(1-x)+\bar{\kappa}_{max} x, \quad\quad x \in (0,1).
\end{equation}

\begin{figure}[ht]
\centering
\includegraphics[width=\linewidth]{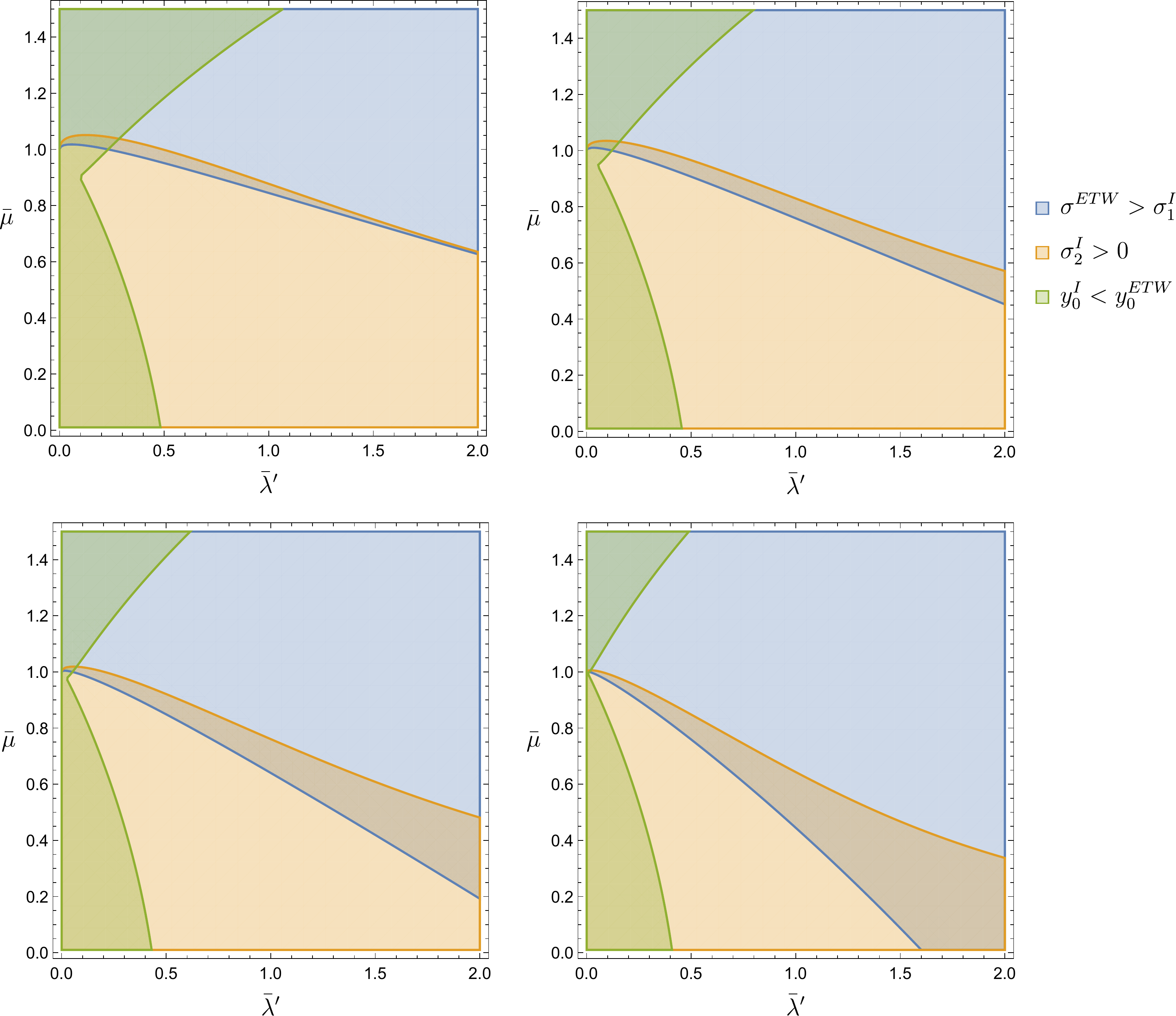} 

\caption{Plots of the regions in the parameter space $(\bar{\lambda}',\bar{\mu})$ with $\sigma^{ETW}>\sigma^I_1$, $\sigma^I_2>0$, or $y^I_0<y^{ETW}_0$ with $d=4$, for the sub-critical value $\bar T=0.6 < \bar T_{*}$. Reading from top left to bottom right: $x=0.16,0.12,0.08,0.04$. We see that there is a small region where all constraints are satisfied. As we decrease $x$, the constraint that $y^I_0 < y^{ETW}_0$ becomes more constraining -- this is because, as noted earlier, the interface brane moves towards the boundary as we approach the lower bound on $\kappa$.}
\label{subcrit}
\end{figure}

\begin{figure}[ht]
\centering
    \includegraphics[width=\linewidth]{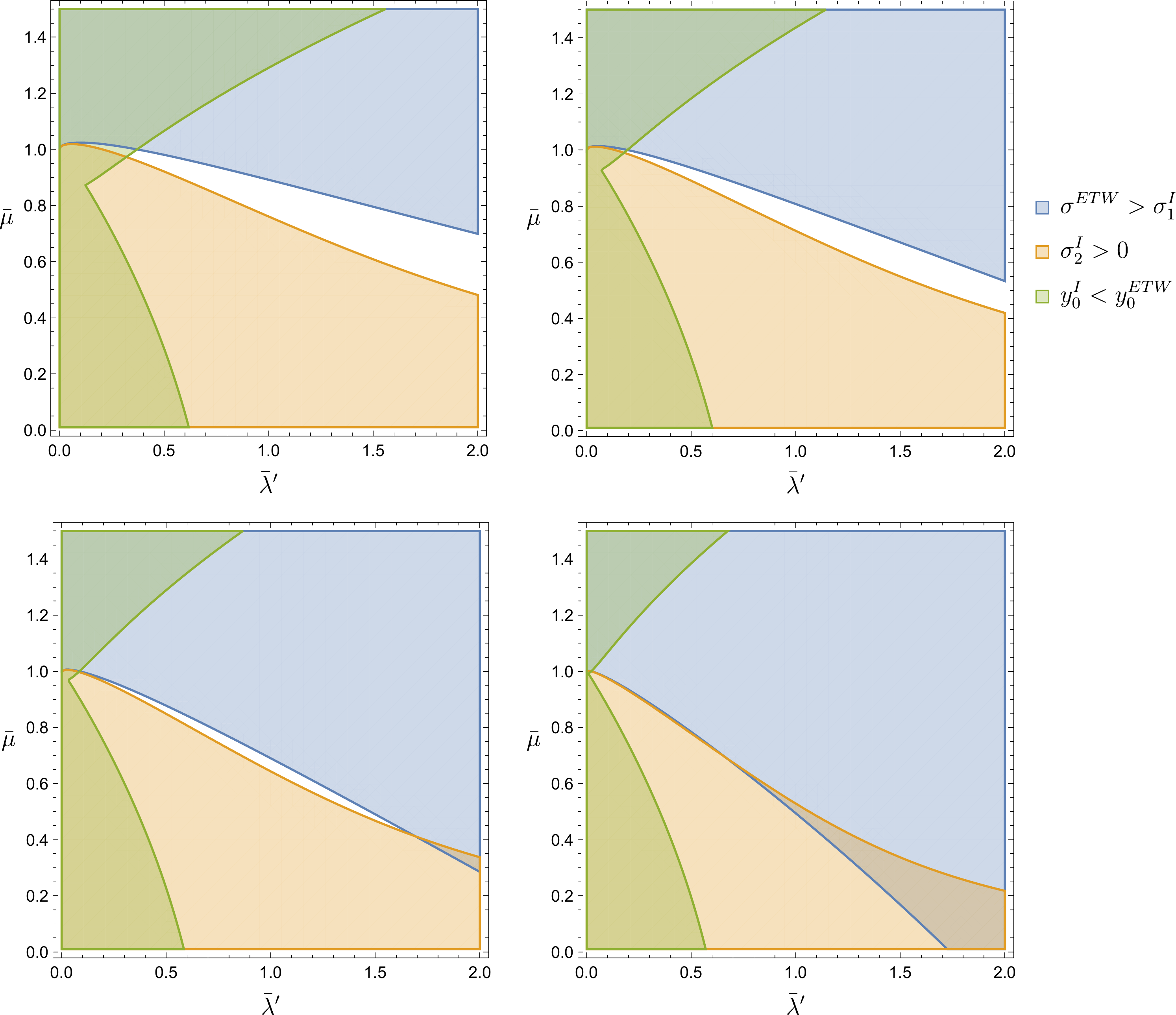}  
\caption{Plots of the regions in the parameter space $(\bar{\lambda}',\bar{\mu})$ with $\sigma^{ETW}>\sigma^I_1$, $\sigma^I_2>0$, or $y^I_0<y^{ETW}_0$ with $d=4$, $\bar T= \bar T_{*}\approx 0.7977$. Reading from top left to bottom right: $x=0.08,0.06,0.04,0.02$. At this critical value, we see that the only point where all the constraints are satisfied is $\bar \lambda'=0$, $\bar \mu =1$, for all values of $x$. This corresponds to the solution with no interface brane, as $\bar \kappa_{min} = \bar \kappa_{max} =0$ for $\bar \lambda'=0$, $\bar \mu =1$.}
\label{crit}
\end{figure}

We plot the allowed regions as functions of $\bar \lambda', \bar \mu$ for a range of values of $\bar T$, $x$. In figure \ref{subcrit}, we plot this for a value $\bar T < \bar T_*$, where we see that there is a small allowed range of values of the parameters where all the constraints are satisfied. In figure \ref{crit}, we plot for $\bar T = \bar T_*$ (to numerical accuracy), seeing that the allowed region shrinks to the single point at $\bar \lambda'=0$, $\bar \mu =1$, for all values of $x$. This corresponds to the solution with no interface brane, as $\bar \kappa_{min} = \bar \kappa_{max} =0$ for $\bar \lambda'=0$, $\bar \mu =1$.  

Thus, while there are allowed solutions with interface branes, these only exist for choices of $T$ that also admit solutions without interface branes, and the range of possible values for the interface brane tension shrinks to zero as the critical value is reached. Allowing the possibility of interface branes has not enlarged the range of possible ETW branes for which we can find eternal wormholes, or good Euclidean solutions describing the initial conditions for our cosmology.

\section{Discussion} 
\label{disc}

In this paper we considered the construction of holographic models for studying closed FRW cosmologies using ETW branes, following \cite{Cooper:2018cmb}. An issue with the construction of such models for $d>2$ is that the requirements of having a good Euclidean solution and having a separation of scales which gives us Einstein gravity on the ETW brane are difficult to satisfy simultaneously. 

We studied a proposed model for satisfying these conditions using interface branes from \cite{VanRaamsdonk:2021qgv}. We found that the use of interface branes relaxes the constraint in the simple model, but introduces new constraints, from requiring that the ETW brane and the interface brane don't collide. Taking into account all the constraints, we find that adding interface branes doesn't increase the range of values of the ETW brane tension for which we have good solutions. 

A key question is whether this failure is essentially technical, or is indicative of a deeper issue. In \cite{VanRaamsdonk:2021qgv}, it was observed that the Euclidean solutions we are trying to construct are related to eternal traversable wormhole solutions, and it has previously been found \cite{Freivogel:2019lej} that it is difficult to construct such solutions in higher dimensions, which might suggest that the problem is deeper. However, we are not aware of any clear physical obstruction to the existence of such  eternal traversable wormhole solutions either; see also the discussion of (non-eternal) traversable wormholes in \cite{Harlow:2021dfp}. 

One hint that the connection to eternal traversable wormholes is relevant to the obstruction we find is that the known way around the obstruction  \cite{Antonini:2019qkt} does not give a wormhole solution. By replacing the Schwarzschild-AdS solution we have considered by a charged black hole, we can decouple the period $\beta$ and the horizon scale $r_h$. By taking the black hole close to extremality we can make $\beta$ arbitrarily large, allowing us to keep $\tau^{ETW}$ in \eqref{tetw} positive as $r_0^{ETW}$ grows \cite{Antonini:2019qkt}. However, if we have a real electric or magnetic charge in the Lorentzian black hole solution, after we analytically continue $t$ and one of the $x^a$, the field will be imaginary, so we do not obtain a good wormhole solution. If we analytically continue the charge parameter, the analytically continued solution does not have an extremal limit, so we can't make $\beta$ large.  It would be very interesting to find some other method for constructing eternal traversable wormholes or brane cosmologies in $d>2$ to explore these questions further. 

\section*{Acknowledgements}

SFR is supported in part by STFC through grant ST/T000708/1, and SF is supported by an STFC studentship.

\bibliographystyle{utphys}
\bibliography{bib}

\end{document}